# SEISMIC COLLAPSE PREDICTION OF FRAME STRUCTURES BY MEANS OF GENETIC ALGORITHMS

A. Greco[a], F. Cannizzaro[a], A. Pluchino[b]

[a]Department of Civil Engineering and Architecture,
[b]Department of Physics and Astronomy
University of Catania, viale A. Doria 6, Catania, Italy
email: agreco@dica.unict.it,
francesco.cannizzaro@dica.unict.it, alessandro@pluchino.it,

## Abstract

This paper presents an automatic approach for the evaluation of the plastic load and failure modes of planar frames. The method is based on the generation of elementary collapse mechanisms and on their linear combination aimed at minimizing the collapse load factor. The minimization procedure is efficiently performed by means of genetic algorithms which allow to compute an approximate collapse load factor, and the correspondent failure mode, with sufficient accuracy in a very short computing time. A user-friendly original software in the agent-based programming language Netlogo, here employed for the first time with structural engineering purposes, has been developed showing its great versatility and advantages. Many applications have been performed both with reference to the classical plastic analysis approach, in which all the loads increase proportionally, and with a seismic point of view considering a system of horizontal forces whose magnitude increases while the vertical loads are assumed to be constant. In this latter case a parametric study has been performed aiming at evaluating the influence of some geometric, mechanical and load distribution parameters on the ultimate collapse load of planar frames.

**Corresponding author**: A.Greco, e-mail: *agreco@dica.unict.it*, tel: +390957382251

# 1 Introduction

The problem of plastic analysis and design of frame structures has been deeply analysed by many researchers since the middle of the past century [1] by means of two different approaches. The first of these is the finite element method in which the global stiffness matrix of the system is computed and the response of the structure is obtained solving iteratively a set of non linear equations [2]. In this analysis the history of loading is applied incrementally until the failure of the structure, therefore the analysis can be very time-consuming.

The second approach is directly based on the kinematic theorem of limit analysis; by analysing all the possible collapse mechanisms of a structure and the related collapse loads, the correct ultimate load is determined seeking the absolute lowest value among the considered mechanisms. This method therefore does not require the direct computation of stiffness matrix and it is not necessary to apply the complete history of loading.

Many contributions can be found in the literature aiming at solving plastic collapse problems for example by means of linear programming [3-5] or at the optimal plastic design of beams [6] or frames considering as constraint the minimum weight of the structure [7-10] or a generic cost function of design variables [11].

In the limit analysis kinematic approach, one of the most frequently used method is that first developed by Neal and Symonds [8,9] in which only the elementary mechanisms are analysed and these are combined to obtain a final collapse mechanism whose load factor is lower than all the possible combinations. The mechanism associated to the lowest load factor represents the real failure mechanism of the structure.

In the literature significant contributions to the automatic computation of the mechanisms for limit analysis of frames have been given by Watwood [10], Gorman [11], Deeks [12] and Kaveh [13].

The main limitation of plastic analysis and design of frames using a combination of elementary mechanisms is the tedious work of combining them to find the true collapse mechanism. Since both steps of generating the elementary mechanisms and combining them are time-consuming, it is therefore important to develop a methodology capable of finding an approximate collapse load factor and the corresponding mechanism as fast and accurate as possible. To this purpose many optimization procedures have been presented in the literature in the last decades. Among the others very interesting approaches may be found in heuristic algorithms such as neural networks, genetic and ant colony algorithms which have the capability to converge on a good solution independently on the specific search space to which they are applied. These algorithms can be used for many optimization problems of different nature.

The present study is devoted to genetic algorithms whose logic and structure are well described in the literature [14,15]. Kaveh et al. [16,17] and Kohama et al. [18] presented studies on the determination of collapse load factors of planar frames by means of genetic algorithms. Kaveh and Jahanshahi studied the plastic limit analysis of frames using heuristic algorithms and ant colony systems [19-21]. A comparison between the performance of co-evolutionary and genetic algorithms in the optimal design of structures has been presented by Hofmeyer et al. [22]. Rafiq proposed a structured genetic algorithm for the optimum design of buildings [23]. A combined parametric modelling and genetic algorithm for performance-oriented process in structural design has been proposed by Turrin et al. [24]. Aminian et al. developed a hybrid genetic and simulated annealing method for estimating base shear of plane structures [25] and also studied the collapse of castellated beams [26]. Recently Jahanshahi et al. [27]

proposed a comparative study on the determination of collapse load of planar frames by means of neural networks, genetic and ant colony algorithms.

The present paper deals with the already mentioned strategy of seeking the collapse load by means of genetic algorithms. The main novelty of the paper consists on the automatic computation of elementary collapse mechanisms of planar frames, based on a regular grid, by means of an original software code in the agent-based programming language Netlogo [28]. This software environment has been usually adopted by many researchers to simulate complex systems dynamics in several different fields, but, at the authors' knowledge, never to structural engineering. In this context, the developed code, together with the correspondent user interface, allows visualizing every single mechanism and the correspondent collapse load, thus providing also a very useful learning tool. Successively, in a very small computing time, the elementary mechanisms are combined and the minimum collapse load is obtained by means of an optimization procedure based on genetic algorithms. The peculiar aspects of the application of genetic algorithms to collapse mechanisms have been described in the paper together with the fundamentals of Netlogo's logic.

Several applications have been performed both with reference to the classical plastic analysis approach, in which all the loads increase proportionally, and with a seismic point of view considering a system of horizontal forces whose magnitude increases while the vertical loads are assumed to be constant.

While the aim of the first applications is mainly to validate the proposed approach by comparison with some of the available results provided in the literature and to study the performance of genetic algorithms, the seismic applications represent an original contribution towards the limit behaviour of structures under earthquake excitations.

In case of seismic analysis the plastic hinges may occur at the two ends of the columns and, due to permanent vertical loads, also in any section along the beam. For this reason a correct location of the plastic hinges along the beam is required as proposed by Mazzolani et al [29].

In the present paper the values of the collapse load, obtained by means of the proposed method for seismic applications, have been compared to the correspondent results provided by nonlinear push over analysis showing a very good correspondence.

Furthermore, an extensive parametric study has been performed aiming at evaluating the influence of some geometric and mechanical parameters, of the intensity of permanent vertical weights, and of the shape of the horizontal force distribution on the ultimate collapse load of planar frames.

The achieved results, with reference to the parametric studies, not only may provide significant information on the seismic performance of frame structures, but also represent a useful tool in their optimal design. In fact, although the presented results refer to some specific structures, general trends in the seismic behaviour of planar frames can be deduced.

## 2   Combination of elementary mechanisms and collapse load

In the present study planar regular frames are considered. These are characterized by the number of floors $N_f$ and the number of columns $N_c$.

The columns at the ground level are assumed to be clamped, therefore the degree of hyperstaticity of the structure, denoted by $h$, is:

$$h = 3\left[N_c - 1 + \left(N_f - 1\right)\left(N_c - 1\right)\right] \qquad (1)$$

By considering the $i$-th floor and the $j$-th column of the frame, the plastic moments of the structural members are assumed to be $M_{b,ij}$ for beams and $M_{c,ij}$ for columns.

The frame may be loaded, at each floor, by concentrated horizontal $F_{h,i}$ and either by vertical incremental forces $F_{v,ij}$, applied in the mid span, or by permanent vertical distributed loads $q_{ij}$. Plastic hinges in the collapse mechanism can therefore be located in $s$ "critical sections" correspondent to each joint and to a certain section of each beam which can coincide with its middle, in case of concentrated forces, or it can vary along the span in case of distributed loading. It is worth to point out that in the case of more than two members converging in a joint, a different critical section must be considered for each member (Figure 1).

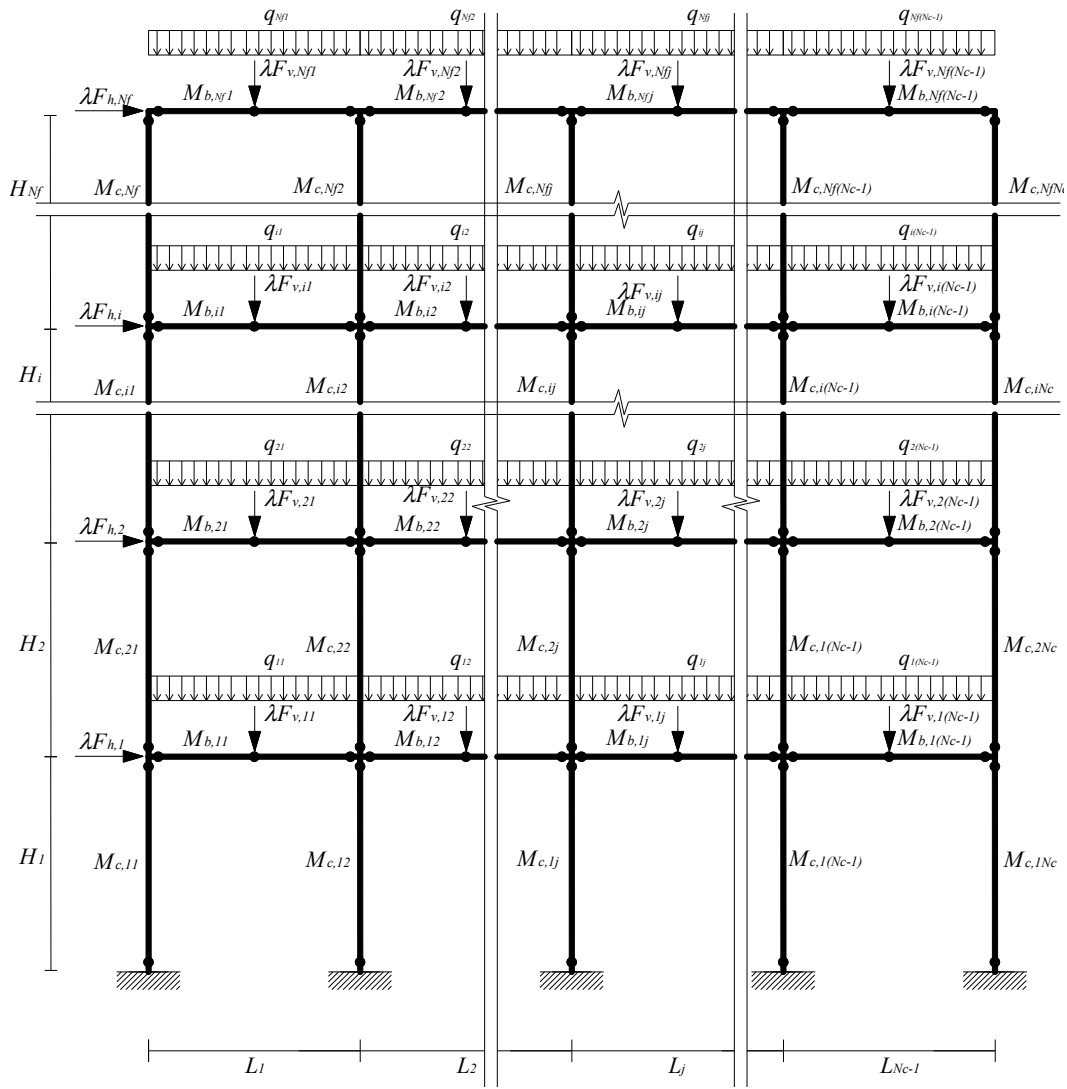

Figure 1. Layout of a generic planar frame

In a kinematic plastic analysis approach the total number of possible collapse mechanism should be considered. Nevertheless, as introduced by Neal and Symonds [8,9] only a small number $e$ of independent elementary collapse mechanisms can be taken into account. Denoting with $s$ the total number of critical sections the following relation holds:

$$e = s - h \tag{2}$$

In the present paper, following the approach of Neal and Symonds [8,9], three different elementary collapse mechanisms are considered: floor, beam and node mechanisms.

In Figure 2 the three independent elementary collapse mechanisms for a generic frame are reported for clarity.

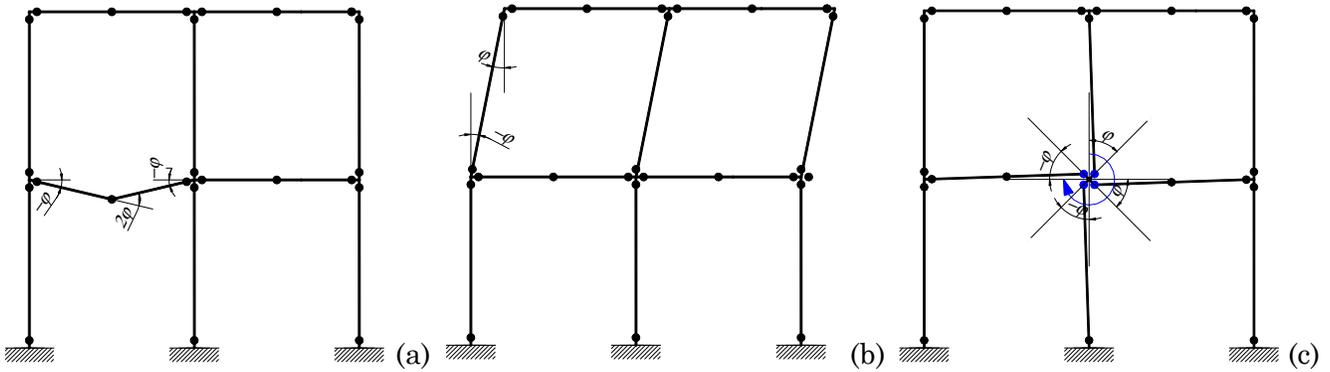

Figure 2. Typical elementary mechanisms: (a) beam mechanism, (b) floor mechanism and (c) node mechanism.

Denoting with $N_{fm}$, $N_{bm}$, $N_{nm}$ respectively the numbers of floor, beam and node mechanism the following relations hold:

$$\begin{aligned} e &= N_{fm} + N_{bm} + N_{nm} \\ N_{fm} &= N_f \\ N_{bm} &= N_f \left( N_c - 1 \right) \\ N_{nm} &= N_c \left( N_f - 1 \right) + N_c - 2 \end{aligned} \tag{3}$$

As it is very well known, according to this method, the elementary mechanisms must be combined each other in order to give the lowest collapse load.

*2.1 Proportional load*

Under the hypothesis that all the applied loads increase proportionally, for each collapse mechanism, either elementary or combined, the virtual work theorem states:

$$\lambda_c W_{ext} = W_{int} \tag{4}$$

where the external virtual work $W_{ext}$ is done by all the non-amplified vertical and horizontal concentrated forces $F_{h,i}$ and $F_{v,ij}$ for the correspondent displacements $d_{h,i}$ and $d_{v,ij}$, while the internal virtual work is referred to the plastic hinges and is the sum of the products of the plastic moments of beams $M_{pb,ij}$ and columns $M_{pc,ij}$ for the related rotations.

As shown in figure 2, for each floor and node elementary mechanism, all the rotations have the same value (but may differ in sign according to the stretched fibers), while for beam mechanism the rotations at the ends of the element are the same and the one at the centre is double with opposite sign. Therefore denoting with $\varphi$ the value of the common rotations, the internal work (dissipated energy) for each elementary mechanism is given by:

$$W_{int}(beam\ mech)_{ij} = 4M_{b,ij}\varphi$$

$$W_{int}(floor\ mech)_i = \sum_{j=1}^{N_C} 2M_{c,ij}\varphi \tag{5}$$

$$W_{int}(node\ mech)_{ij} = \left(M_{b,(i-1)j} + M_{b,ij} + M_{c,ij} + M_{c,i(j+1)}\right)\varphi$$

where the terms in the node mechanisms can be whether considered or not according to the presence of the corresponding members converging in the node.

The external work related to node mechanism is zero while, under the hypothesis of small rigid rotations, it is:

$$W_{ext}(beam\ mech)_{ij} = F_{v,ij}d_{v,ij} = 0.5F_{v,ij}L_j\varphi$$

$$W_{ext}(floor\ mech)_i = \left(\sum_{k=i}^{N_f} F_{h,k}\right)d_{h,i} = \left(\sum_{k=i}^{N_f} F_{h,k}\right)H_i\varphi \tag{6}$$

When the elementary mechanisms are combined, the rotations related to each critical section are simply obtained adding all the relevant values. The external work in a combined mechanism is the sum of the elementary ones while the internal work is computed by multiplying each total rotation for the plastic moment of the related element. Therefore, when elementary mechanisms with opposite rotations in a critical section are added, the total rotation may turn out to be zero thus providing a smaller dissipated energy. Of course in order to obtain the lowest value of $\lambda_c$, the dissipated energy must be as smaller as possible, and this is fulfilled by means of the optimization procedure.

*2.2 Seismic load*

In the case in which only the horizontal forces are considered variable, as in the case of seismic analysis of structures, while the vertical loads are assumed to be distributed and of constant value, for each collapse mechanism the virtual work theorem states:

$$\lambda_c W_{ext} + W_{extV} = W_{int} \tag{7}$$

where $W_{extV}$ represents the work done by the vertical permanent load, which are not magnified. Therefore the value of the multiplier $\lambda_c$ in this case is given by:

$$\lambda_c = \frac{W_{int} - W_{extV}}{W_{ext}} \tag{8}$$

With reference to the external work, different distributions of the horizontal forces $F_{h,i}$ considered applied at each floor level can be selected according to a fixed shape, while distributed constant vertical loads $q_{ij}$ act on each of the beams.

The external work done by the permanent vertical load for each beam is related to the beam mechanism shown in Figure 3 where a plastic hinge can be opened at a position $x_{ij}$ from the left end of the beam dependent on the magnitude of the uniform load acting on the beam

In fact, at any loading stage the bending moment in each beam is the superposition of the one due to the uniformly distributed vertical loads and that due to horizontal forces. Therefore, increasing the horizontal forces the first plastic hinge opens at the beam end opposite to the horizontal force, with plastic moment $M_{b,ij}$, while the second hinge forms at the first end when the distributed load is equal to the limit value [27]:

$$q_{\lim,ij} = \frac{4M_{b,ij}}{L_j^2} \qquad (9)$$

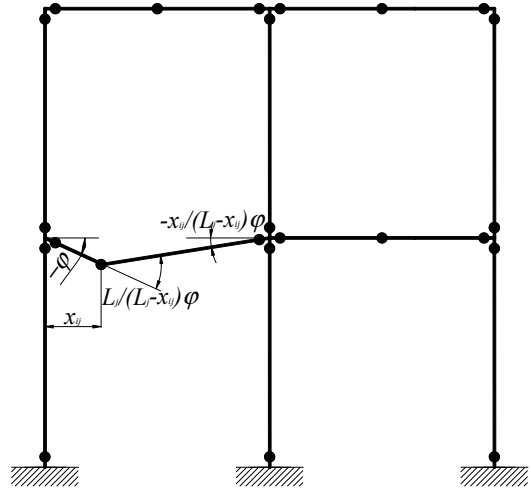

Figure 3. Beam mechanism in case of permanent vertical loads

When the vertical load $q_{ij}$ exceeds the limit value provided in (9) a hinge can open along the beam at the abscissa

$$x_{ij} = L_j - 2\sqrt{\frac{M_{b,ij}}{q_{ij}}} \qquad (10)$$

Therefore, the external work for each beam and floor mechanism is:

$$W_{ext}(beam\ mech)_{ij} = 0.5 q_{ij} x_{ij} L_j \varphi$$

$$W_{ext}(floor\ mech)_i = \left(\sum_{k=i}^{N_f} F_{h,k}\right) d_{h,i} = \left(\sum_{k=i}^{N_f} F_{h,k}\right) H_i \varphi \qquad (11)$$

For the internal work related to floor and node elementary mechanisms the same expressions of the case of proportional load hold, while for beam mechanisms it is (Figure 3):

$$W_{int}(beam\ mech)_{ij} = 2M_{b,ij}\left(1 + \frac{x_{ij}}{L_j - x_{ij}}\right)\varphi \qquad (12)$$

For the evaluation of the external and internal work in a combined mechanism, the considerations described in the case of proportional load hold.

## 3  Optimization procedure through genetic algorithm

As it is very well known, analyzing all the possible combinations of elementary mechanisms, the minimum value of $\lambda_c$ must be sought in order to obtain the real collapse load.

Many studies have been presented in the literature dealing with the method of combining elementary mechanisms proposed by Neal and Symonds [8,9]. In this paragraph the optimization procedure adopted in the present paper, which makes use of genetic algorithms, is described.

A "genetic algorithm" is an adaptive stochastic method that mimic the Darwinian evolution, based on an opportune combination of random mutations and natural selection, in order to numerically find optimal values of some specific function. The algorithm acts over a population of $P$ potential solutions by applying, iteratively, the "survival of the fittest" principle: in such a way it produces a sequence of new generations of individuals that evolves towards a stationary population where the large majority of surviving solutions do coincide and approach as much as possible the real solution of a practical problem [15].

In order to translate into this scenario the original problem of finding the linear combination of elementary collapse ss with the minimum value of $\lambda_c$, corresponding to the true collapse load factor, each individual of the population is coded as a string of integer numbers, called chromosome, where each number (called gene) represents how many times a given elementary mechanism enters in the combination. Therefore, if there are in total $N = N_{fm} + N_{bm} + N_{nm}$ elementary mechanisms, a generic chromosome $C_i$ of the population ($i = 1, ..., P$) represents a generic weighted combination of those mechanisms and can be coded in the following string:

$$C_i \quad (c_1, c_2, c_3, ... c_k, ..., c_N) \qquad (13)$$

with $c_k \in [0, c_{max}]$, being $c_{max}$ the maximum number of times the $k$-th mechanism is involved in the combination ($c_k = 0$ means that the mechanism is not involved at all). Given the string, it is possible to calculate the load factor $\lambda_i$ of the corresponding combination (chromosome). The overall number of possible different chromosomes is thus $P_{max} = (c_{max} + 1)^N$, a quantity which rapidly increases with $N$ even for small values of $c_{max}$ (for example, if $c_{max} = 2$ and $N = 22$, one obtains $P_{max} \approx 31 \cdot 10^9$). Task of the genetic algorithm is that of exploring the space of all the possible chromosomes, in search of the combination of mechanisms which minimizes $\lambda_i$ or, that is the same, which maximizes an opportune "fitness function", here defined as

$$f(\lambda_i) = K - \lambda_i \qquad (14)$$

where $K$ is an arbitrary constant, chosen great enough to have $f_i > 0$ for every possible value of $\lambda_c$. In the following it will be set $K = 100$ without loss of generality. The fitness value $f(\lambda_i)$, associated to each chromosome, represents the probability of survival of that chromosome, under the pressure of the natural selection process. In the following it is explained more in detail how this process does work.

Starting from the initial population of $P$ chromosomes (typically $P = 100$), randomly chosen among the $P_{max}$, a new generation is created from the old one, where chromosomes that have a higher fitness score are more likely to be chosen as "parent" than those that have low fitness scores. The selection method adopted in this paper is called "tournament selection", with a tournament size of three: this means that groups of 3 chromosomes are drawn randomly from the old generation, and the one with the highest fitness in each group is chosen to become a parent. Either one or two parents are chosen to create children: with one parent, the child is simply a clone of the parent; with two parents, the process is the digital analog of sexual recombination – the two children inherit part of their genetic material from one parent and part from the other (crossing-over). Once the new generation is created, there is also a chance that random mutations will occur at level of the single genes $c_k$ of the child chromosomes, and some of them will be changed into new ones (always chosen in the interval $[0,c_{max}]$). In Figure 4 a sketch of these three operations is summarized.

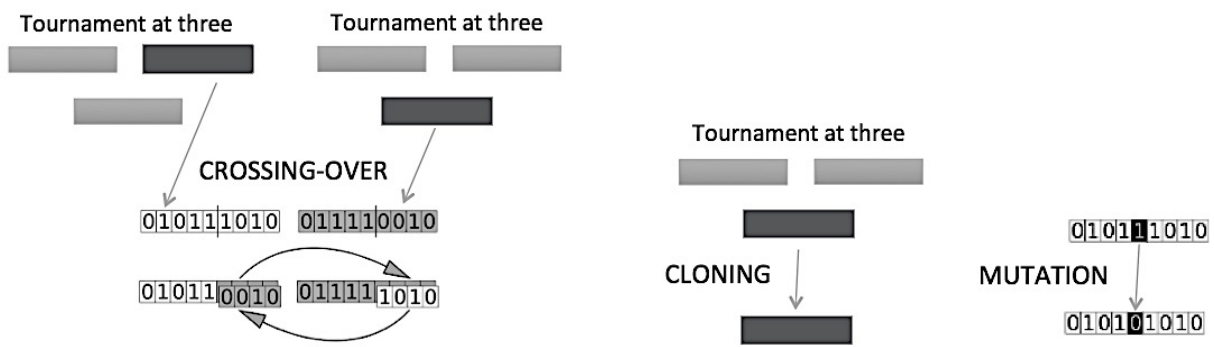

Figure 4. Natural selection process through tournament with size of three. Boxes in dark grey represent chromosomes with higher fitness. In this example $c_{max} = 1$, i.e. binary genes are considered.

By iteratively repeating this process several times, chromosomes with the highest fitness will be progressively selected in the space of all the possible combinations and will quickly spread among the population reducing the diversity of the individuals, until (almost) only one of them will survive: hopefully, that one with the maximum fitness (and, correspondingly, with the minimum load factor). Of course, it is frequent for the dynamics to remain trapped into local maxima of fitness (minima of $\lambda_i$), therefore it is convenient to launch the genetic algorithm many times (events), each time starting from a different initial population, in order to gain more chances to reach the global maximum of fitness, corresponding to the true collapse load factor of the structure.

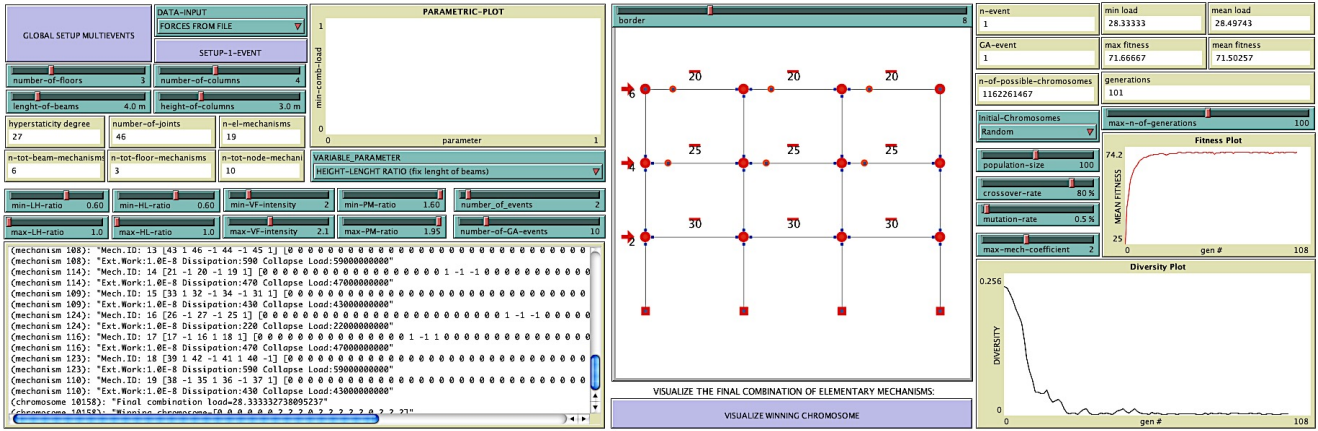

Figure 5. Graphical aspect of the user interface of our program in the NetLogo environment.

*3.1 The proposed NetLogo code*

One of the main goals of this paper is to develop an original code for both the calculation ab origine of all the elementary collapse mechanisms of a given planar frame and the implementation of the genetic algorithm described above for the (fast and reliable) determination of the combination of elementary mechanisms with the minimum collapse load factor for the structure considered. For this purpose a very powerful software has been adopted, that is NetLogo [28], which is a freeware multiplatform environment with an owner high level programming language and with a very ductile and versatile user interface.

NetLogo platform was natively developed for agent based simulations and for modeling complex systems behavior, therefore – as far as the authors know – the present one is the first application to the study of planar frames and collapse mechanisms. The idea is to harness the power of the NetLogo graphical user interface and the versatility of its agent-oriented programming language in order to create a user-friendly original software for the complete and automatic analysis of the limit behaviour of structures under earthquake excitations.

In Figure 5 the final layout of the user interface in the NetLogo environment is reported, as it appears after a single event analysis of a generic frame with three floors and four columns. The interface is visually organized in two parts, corresponding to two conceptually subsequent steps of the analysis: (i) the *central-left* part, dedicated to the input parameters, to the setup of the frame and to the calculation of its elementary mechanisms; the chosen frame is then visualized in the World of NetLogo, which is a two-dimensional square box endowed with a customizable cartesian reference system; (ii) the *right* part, dedicated to the plastic analysis of the frame through the genetic algorithm.

The software elements of each one of these two parts is now described in the following.

(i) The main elements of the *central-left part* of the interface are the two setup buttons, able to launch either a single-event or a multi-event simulation, and the several input sliders needed to set both the parameters of the frame (number of floors and columns, length of beams, height of columns) and the range of variation of the selected variable quantity for the parametric studies (the four options are: *height-length ratio* with fixed length of beams*, length-height ratio* with fixed height of columns, *vertical forces intensity, plastic moments ratio*).

The first time the user inputs a new frame, it is also necessary to manually insert (through a pop-up dialog window) all the supplementary missing input parameters for the various floors of the frame (vertical loads and horizontal forces, beam and column plastic moments); these parameters are stored in an external file and can be retrieved automatically for next simulations with the same frame.

Of course, each setup button call a given subroutine of the code, which contains the instructions needed to build the frame in the World and to calculate the elementary collapse mechanisms. To this aim, the powerful high-level agent-based commands of NetLogo have been exploited, which allow to easily create and simultaneously manipulate many objects (agents) into the World. Three main categories of agents are present in NetLogo: turtles (moving objects), links (edges linking couples of turtles) and patches (active pixels of the World). Beam and columns were represented with links, while nodes, plastic hinges and joints with turtles. The agents of these two categories dynamically interact one with each other inside the World, and all of them also interact with the active patches in order to calculate and visualize the three kind of elementary collapse mechanisms (beam, floor and node ones). In Figure 6 three examples of these mechanisms are shown, as they appear in the World of NetLogo.

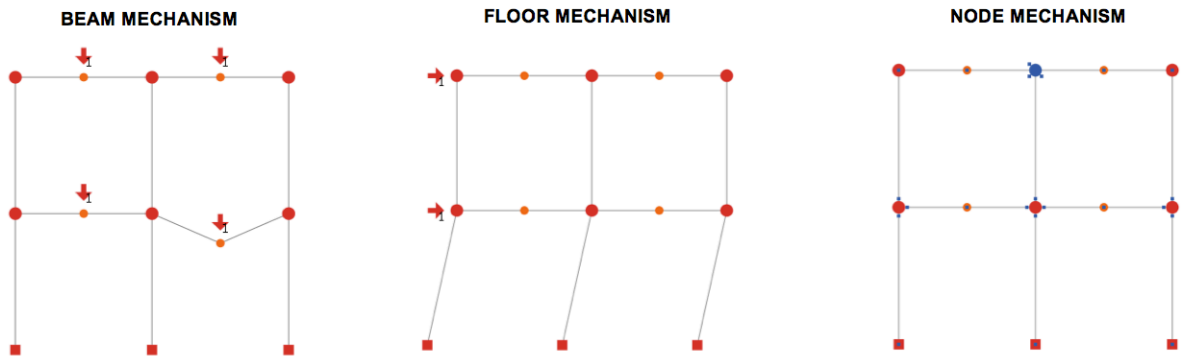

Figure 6. Examples of the three kind of possible elementary collapse mechanisms (beam, floor and node) as they appear in the user interface.

Opportune *monitors* (output boxes) show the total number of elementary mechanisms, the hyperstaticity degree, the number of joints and the partial numbers for the three kind of mechanisms; all of these data are automatically computed.

Once calculated all the elementary mechanisms, together with their collapse load factors, the code is able to aggregate them in linear combinations (chromosomes) and to calculate the corresponding collapse load factors. At this point the genetic algorithm can go into action (as explained in the previous section) in order to find the minimum of those factors which should approximate that of the real collapse mechanism of the plastic frame.

(ii) In the *right part* of the user interface it is possible to set the main parameters of the genetic algorithm, i.e.: the size $P$ of the chromosomes' population, the maximum number of iterations (i.e. of new generations), the cross over rate, the mutation rate and the maximum value ($c_{max}$) for the mechanisms' coefficients.

In particular, the *crossover-rate* slider controls the percentage of each new generation that has to be created through sexual reproduction (recombination or crossing-over between the genes of two parents' chromosomes) and the percentage (*100 – crossover-rate*) to be created through asexual reproduction (cloning of one parent's chromosome), while the mutation-rate slider controls the percentage probability of mutation, which applies to each genes of all the chromosomes in the new generation. As shown later, typical values of these two quantities are about 80% for the crossover rate and about 1% for the mutation rate. On the other hand, typical values for the population size are between 50 and 100 chromosomes, while usually $c_{max}$ is not greater than 3.

In the upper part of the interface several monitors show the static value of the total number of possible different chromosomes, fixed *ab initio* by both $c_{max}$ and the number of elementary mechanisms, and the dynamic values (changing for each new generation) for the mean and minimum collapse load factors of the current population of chromosomes, together with their mean and maximum fitness. The mean fitness and the diversity of the population are also reported as function of the iterations in two different plots. The diversity is evaluated by averaging the Hamming distance (that is the fraction of positions at which two strings have different entries) among all the possible couples of chromosomes in the population.

As visible in Figure 5 for a single-event run, starting from a population of *P = 100* randomly selected chromosomes, 100 iterations are enough for the system to reach a stationary state where a single chromosome, that one with the best fitness and the (local) minimum collapse load, does survive and spreads among the population. At the end of the simulation it is also possible to visualize the combination of elementary mechanisms corresponding to the winning chromosome. On the other hand, as explained in the previous section, for a given set of frame parameters, it is preferable to launch the multi-event version of the algorithm, which runs the same genetic routines several times (as established by the slider *number-of-GA-events*, usually set to 10) with different random realizations of the initial population, in order to escape from local minima of the collapse load and to have more chances to reach a global one.

Finally, the multi-event version can be also used to perform parametric studies of the same frame, by setting the minimum and maximum values of a given variable parameter (height-length ratio, length-height ratio, vertical forces intensity or plastic moments ratio) and by choosing the number of intermediate steps with the slider *number_of_events*.

# 4 Applications

In this section several applications will be presented aiming at validating the proposed approach and performing various parametric studies in order to assess how some of the geometric and mechanical parameters affect the collapse load and mechanism of frame structures.

In the first sub-section the case of frames subjected to a proportionally increasing load distribution is treated to validate the proposed approach on two frames already studied in the literature. For the second example several parametric studies considering the sensitivity of the adopted genetic algorithm, to some characteristic parameters, are performed.

In the second sub-section the case of seismic collapse load is faced with. In this second case a distribution of increasing horizontal loads (seismic contribution) is considered together with a constant vertical load distribution (gravitational loads).

Finally a parametric study is performed in the last subsection. The benchmark frame described in the second sub-section is considered; the sensitivity of the ultimate load is assessed with respect to the ratio between floor height and bay and its reciprocal, to the ratio between plastic moments in beams and columns, with respect to the magnitude of the gravitational loads and with respect to the shape of the distribution of the horizontal loads.

In all the considered cases, the relevant collapse mechanisms are reported and discussed.

*4.1 Plastic collapse for proportionally increasing load*

The first application refers to the three storey frame analyzed by Kaveh and Jahanshahi in [20] and reported in Figure 7 together with the loads and the plastic moments of beams and columns. In the classical hypothesis that all the concentrated forces are uniformly amplified the authors found that the collapse mechanism involves only the first two storeys and is related to a load multiplier equal to 1.6.

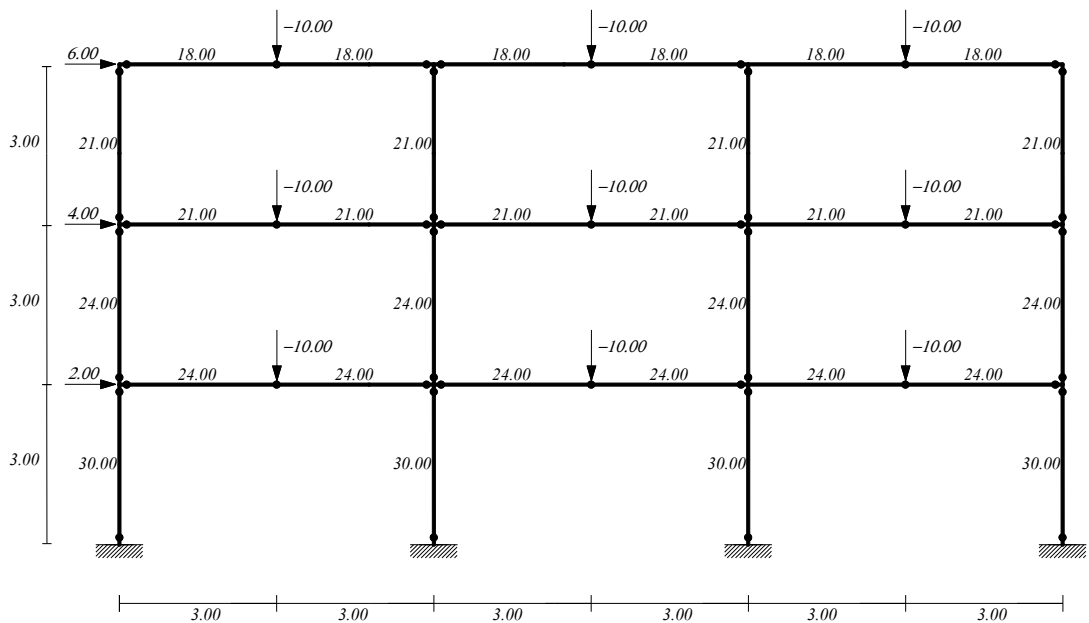

Figure 7. Geometrical and mechanical layout of the first frame studied in [20].

The previously described Netlogo code, setting for the genetic algorithm a population size of 100 individuals, 100 generations with 75% of cross over rate and 1% of mutation rate, provided the load multiplier of $\lambda_c=1.6$ in a computational time of 28.5 sec. The result is the minimum value over 10 runs with a mean value of 1.6105. Figure 8 shows the output of the Netlogo code in which it can be easily noticed that the chromosome correspondent to the minimum load value is related to the same collapse mechanism obtained by Kaveh and Jahanshahi.

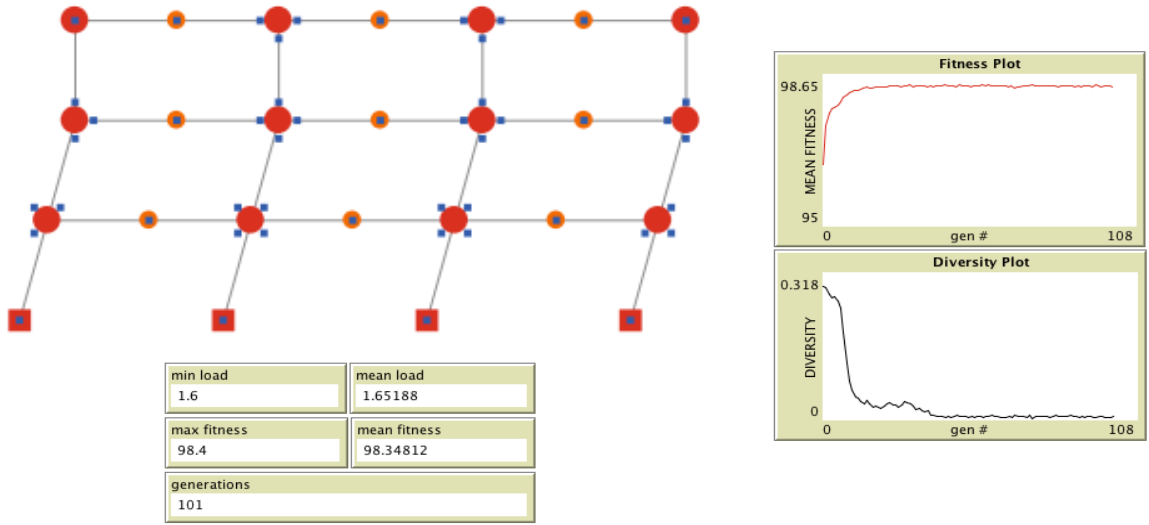

Figure 8. Output of the NetLogo application relative to the frame reported in Figure 7.

Furthermore it is interesting to point out that for this frame the total number of elementary mechanisms (and therefore the length of each chromosome) is equal to 22 and that the total number of possible chromosomes, setting to 3 the maximum coefficient for each mechanism, is 17592186044416. Finally the two plots on the left of the figure show that the fitness increases rapidly with the generation and converges to a value almost equal to 1 while the diversity decreases towards 0.

The second application is referred to the four storey frame also analyzed by Kaveh and Jahanshahi in [20] and reported in Figure 9.

The same load multiplier of 0.65 and collapse mechanism reported by the previous authors are obtained by means of the proposed code as shown in Figure 10. For this frame a sensitivity study to the population size, the percentage of the cross over rate and the mutation rate in the chromosomes of the genetic algorithm has been performed.

The following figures show the results of the minimum value of the collapse load and its mean value over 10 runs when the above mentioned parameters vary in assigned ranges.

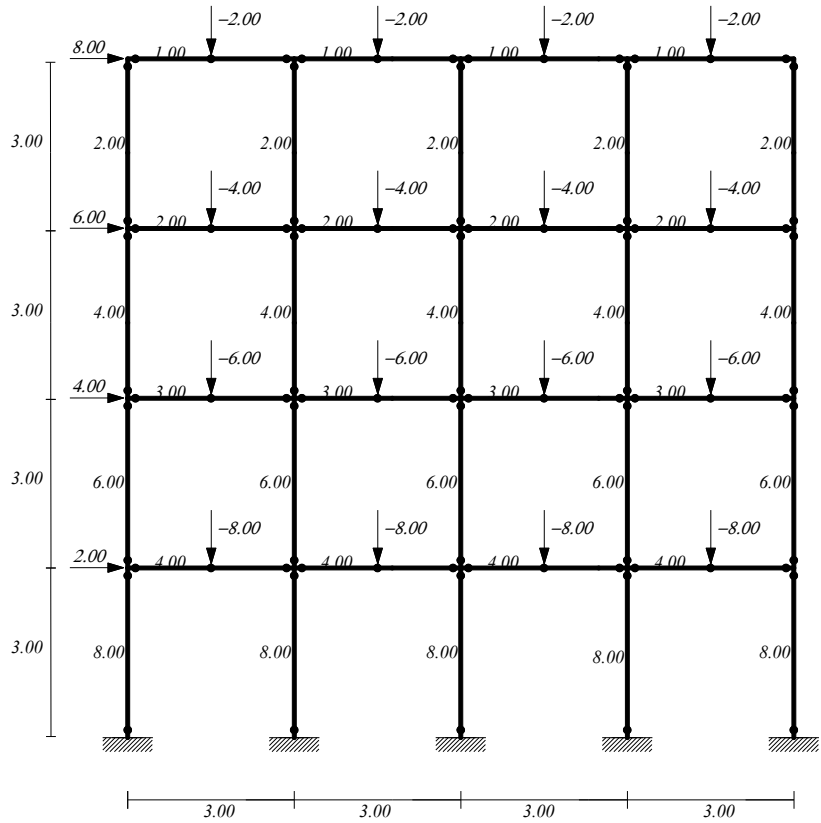

Figure 9. Geometrical and mechanical layout of the second frame studied in [20]

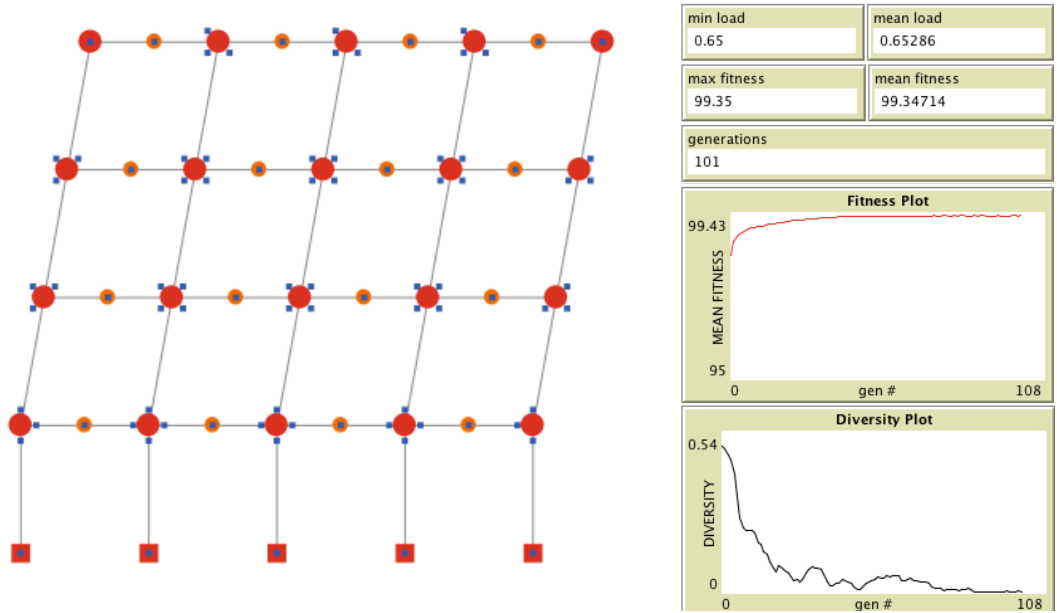

Figure 10. Output of the NetLogo application relative to the frame reported in Figure 9.

In particular, Figure 11 shows that the minimum value of the collapse load is independent on the percentage of cross over rate while its mean value decreases by increasing the cross over rate, reaching its minimum value around 80%. Standard deviation bars show that 80% of cross over rate also ensures the maximum number of correct retrievals of the minimum collapse load. Therefore, in the following applications a fixed cross over rate of 80% will be assumed for generating the new populations of chromosomes.

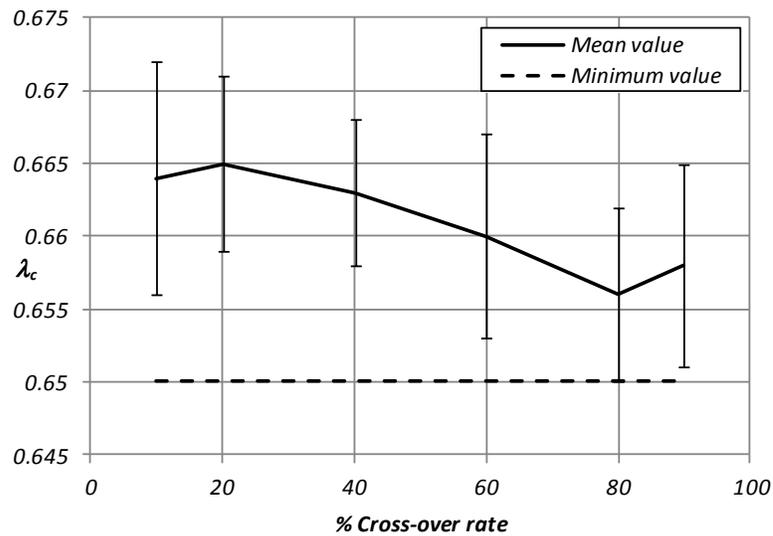

Figure 11. Sensitivity of the cross over rate vs load multiplier

In Figure 12 the sensitivity of the obtained results to the percentage of mutation rate has been shown. As it can be clearly noticed the minimum value of 0,65 for the collapse load has been always determined for mutation rates equal or lower than 3% and then it increases with the percentage of mutation. With reference to the mean value over 10 runs, the best results have been obtained for a mutation rate equal to 1% and therefore this value will be assumed constant in the following applications.

Figure13 shows that an adequate size of the considered population can not be lower than 60. In the following applications reference will be made to populations of 80 chromosomes. In fact, as the population size increases, the minimum value obtained over 10 runs progressively approaches the correct one. When the population size is 40, the correct collapse multiplier is obtained for the first time, but only when the population is equal or higher than 80, the correct collapse multiplier is stably reached.

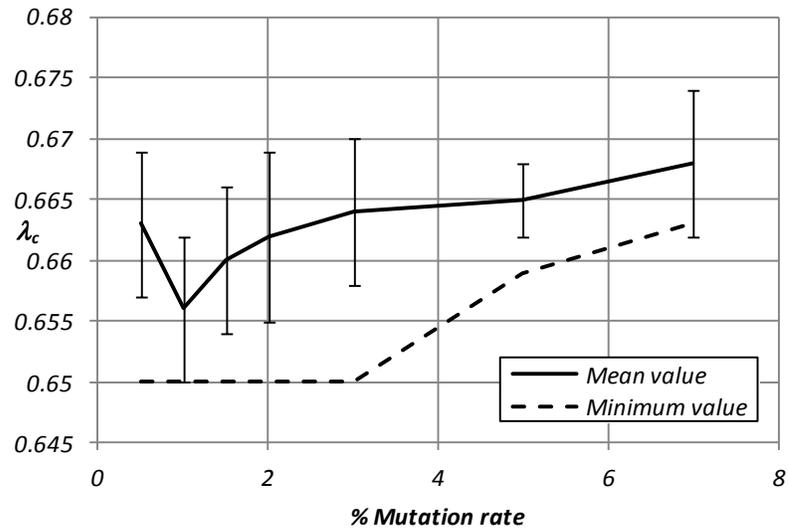

Figure 12. Sensitivity of the mutation rate vs load multiplier

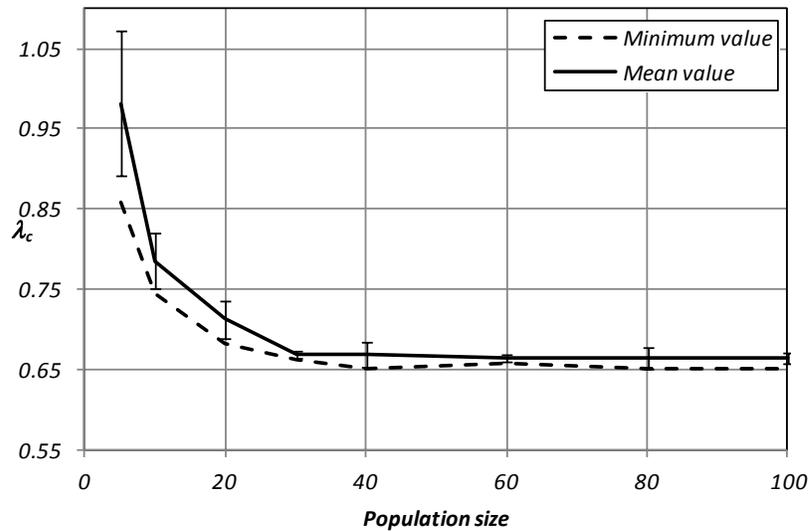

Figure 13. Sensitivity of the population size vs load multiplier

*4.2 Seismic collapse load*

In this section the case of a frame subjected to permanent distributed vertical loads acting on the beams and increasing horizontal floor forces is treated.

The presence of gravity loads may lead to the opening of along axis hinges on the beams when its magnitude is higher than the limit value reported in Eq.(9).

The considered frame is a three storey and three bay one whose geometry is reported in Figure 14. In the same figure the plastic moments are reported next to each beam and column.

The limit vertical distributed loads for each floor, according to Eq.(9), are respectively equal to $q_{\lim,1}=30$, $q_{\lim,2}=22.5$, $q_{\lim,3}=15$, therefore, for the first floor the vertical load acting on the beams is equal to the limit value while for the second and third floors it is greater. In this case the comparison with the proposed procedure has been performed with a classic pushover approach in terms of ultimate load. A numerical model was implemented in the well known FEM software SAP2000 [30]. The implemented model employs lumped nonlinearities, simulated by means of perfectly plastic hinges. Since in a pushover analysis the location of plastic hinges has to be set 'a priori', in the beams the critical sections have been set every 10 cm.

For the considered frame, the FEM approach leads to a collapse multiplier of the horizontal loads equal to $\lambda_{c,SAP}=28.1645$, while with the proposed approach the collapse multiplier is equal to $\lambda_c=28.33$. It is worth to notice that, since the forces are gradually incremented in a pushover analysis, the actual collapse multiplier $\lambda_u$ of a structure represents an upper bound for such an approach. Furthermore, due to the kinematic theorem of plastic analysis $\lambda_u$ represents a lower bound for a limit analysis approach. Therefore it has to be $\lambda_{c,SAP} \leq \lambda_u \leq \lambda_c$.

Figure 15a reports the collapse mechanism for the considered frame and shows that the potential plastic hinges along the span of the beams in the second and third floor do not rotate in the total collapse mechanism. This is in fact characterized by the rotation of the columns at their base and occurrence of plastic hinges at all the ends of the beams.

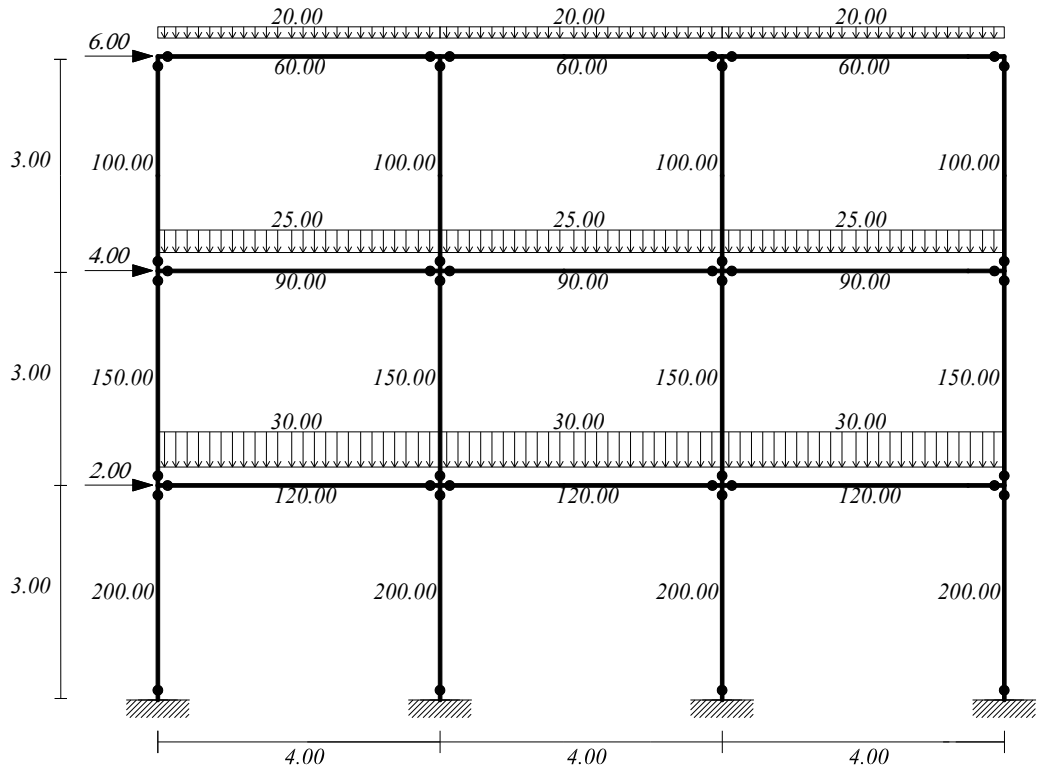

Figure 14. Geometrical and mechanical layout of the frame studied in seismic conditions

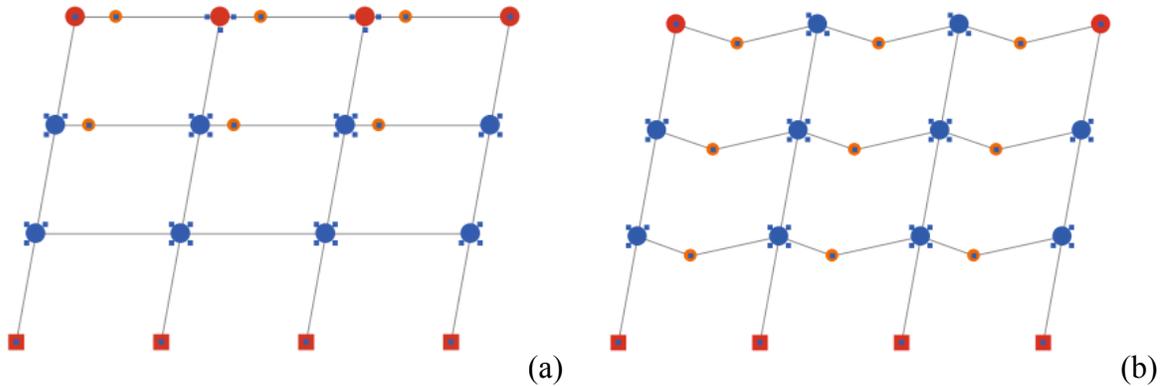

Figure 15. Collapse mechanisms of the frame under seismic conditions with (a) single and (b) double intensity of the vertical loads.

As a further application the case in which all the vertical applied loads assume a double value of the one reported in Figure 14 is analyzed. In this case the collapse multiplier calculated by the proposed approach is equal to $\lambda_c = 24.413$ while SAP2000 provides $\lambda_{c,SAP} = 24.001$.

The collapse mechanism correspondent to the ultimate load multiplier is shown in Figure15b. As it can be clearly observed in this case all the elementary beam mechanisms at each floor are involved in the global collapse mechanism.

*4.3 Parametric study*

Finally, a parametric study to assess the sensitivity of the ultimate load with respect to several parameters is here reported. The benchmark case is that relative to the previous subsection and reported in Figure 14. Five different parameters have been investigated, namely the intensity of the permanent load, the geometric ratio between storey height and bay *H/L* and its reciprocal *L/H*, the ratio between the plastic moments in beams and columns, and the shape of the horizontal loads.

In order to justify some interesting trends in the collapse multiplier, some of the obtained collapse mechanisms, relative to significant values of the investigated parameters, are reported below the pictures and discussed carefully.

In Figure 16 the ultimate load versus a multiplier $\alpha$ of the permanent loads is considered. In particular the conditions $\alpha = 0$ and $\alpha = 1$ represent respectively the cases of no permanent loads acting on beams and permanent load of the benchmark case (Figure 14). The results show a good agreement between the FEM and the proposed approaches. It is worth to notice that as the multiplier $\alpha$ is lower than a transition value equal to 1.06, the permanent load acting on all beams does not imply any plastic rotation located at the hinges along the axis of the beams and the collapse mechanism does not change, therefore the ultimate load is constant in this range. Then, when $\alpha$ increases the ultimate load decreases progressively and plastic hinges also open along the beams producing relative rotations. It can be easily seen, from the collapse mechanisms reported in the same figure, that increasing the value of $\alpha$ the positions of the plastic hinges in the beams move towards the right end.

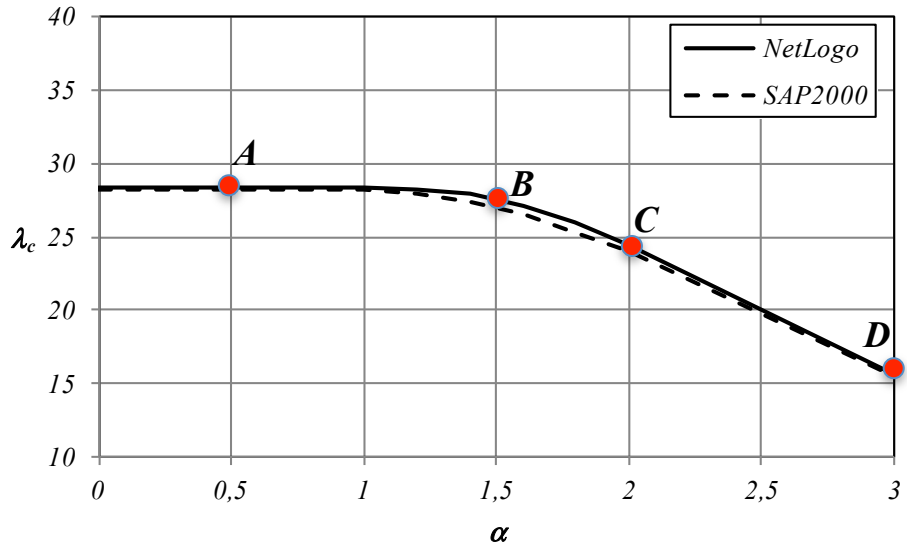
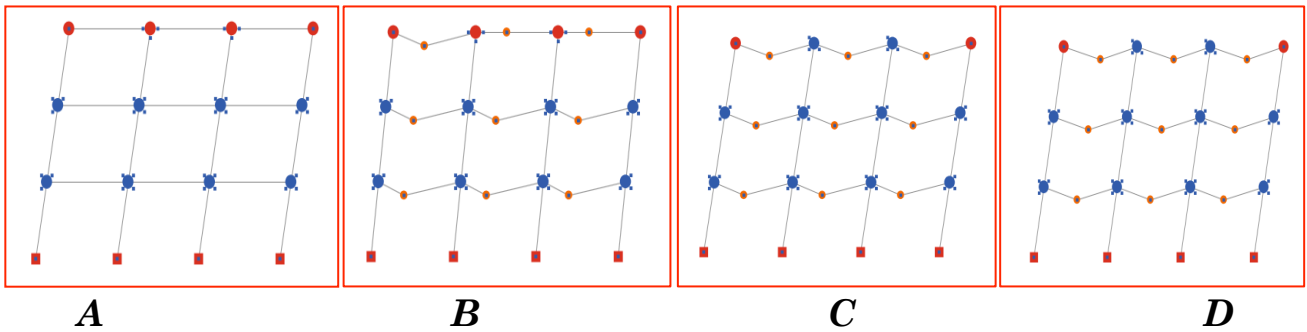

Figure 16. Collapse load multiplier vs intensity of the permanent vertical loads

In the following the results considering a variation of the ratio between the storey height *H* and the bay *L* of the frame are considered. Both the ratios *H/L* (keeping constant the value of 4.00 for the bay) and *L/H* (keeping constant the value of 3.00 for the height of the columns) are taken into account. The comparison between the proposed approach, considering the variation of the ratio *H/L*, and SAP2000 is reported in Figure 17 showing again a very good agreement. As expected for a fixed value of the bay, as the storey height increases the collapse load decreases significantly; as shown in Figure 17 the collapse mechanisms for all the analyzed ratios do not involve beam mechanisms at any storey.

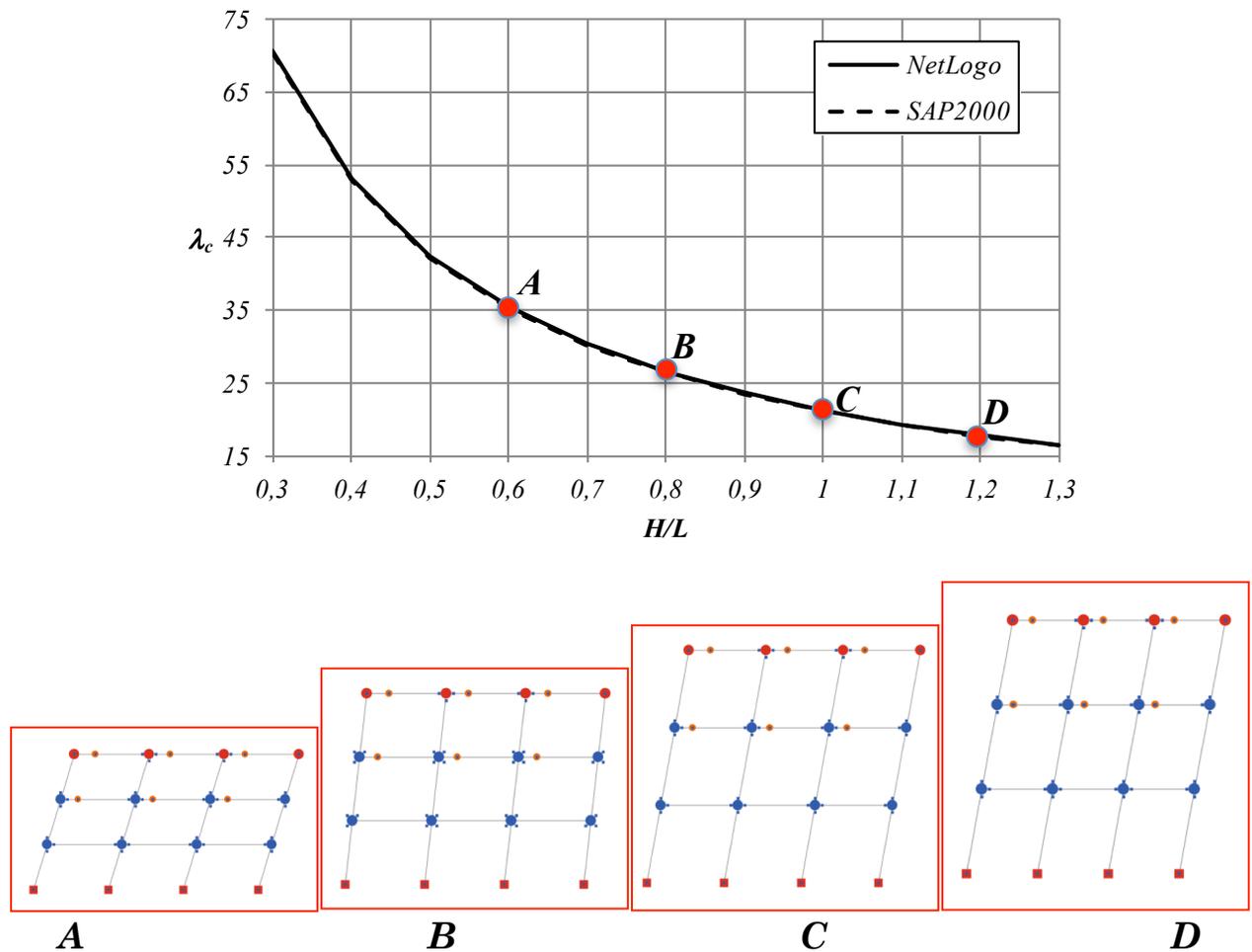

Figure 17. Collapse load multiplier vs ratio between floor height and bay length

When the height of the columns is assumed to be constant and the length of the beams varies from 2.4 (*L/H=0.8*) to 6 (*L/H=2*) the ultimate collapse mechanism changes significantly. In fact, when *L/H* is smaller than 1.38, the frame collapse occurs without the opening of plastic hinges in the beams and the collapse load, referred only to floor mechanisms, is constant. For higher values of the length of the beams the collapse load trend changes abruptly with a progressive decrease and, as shown in Figure 18, beam mechanisms arise.

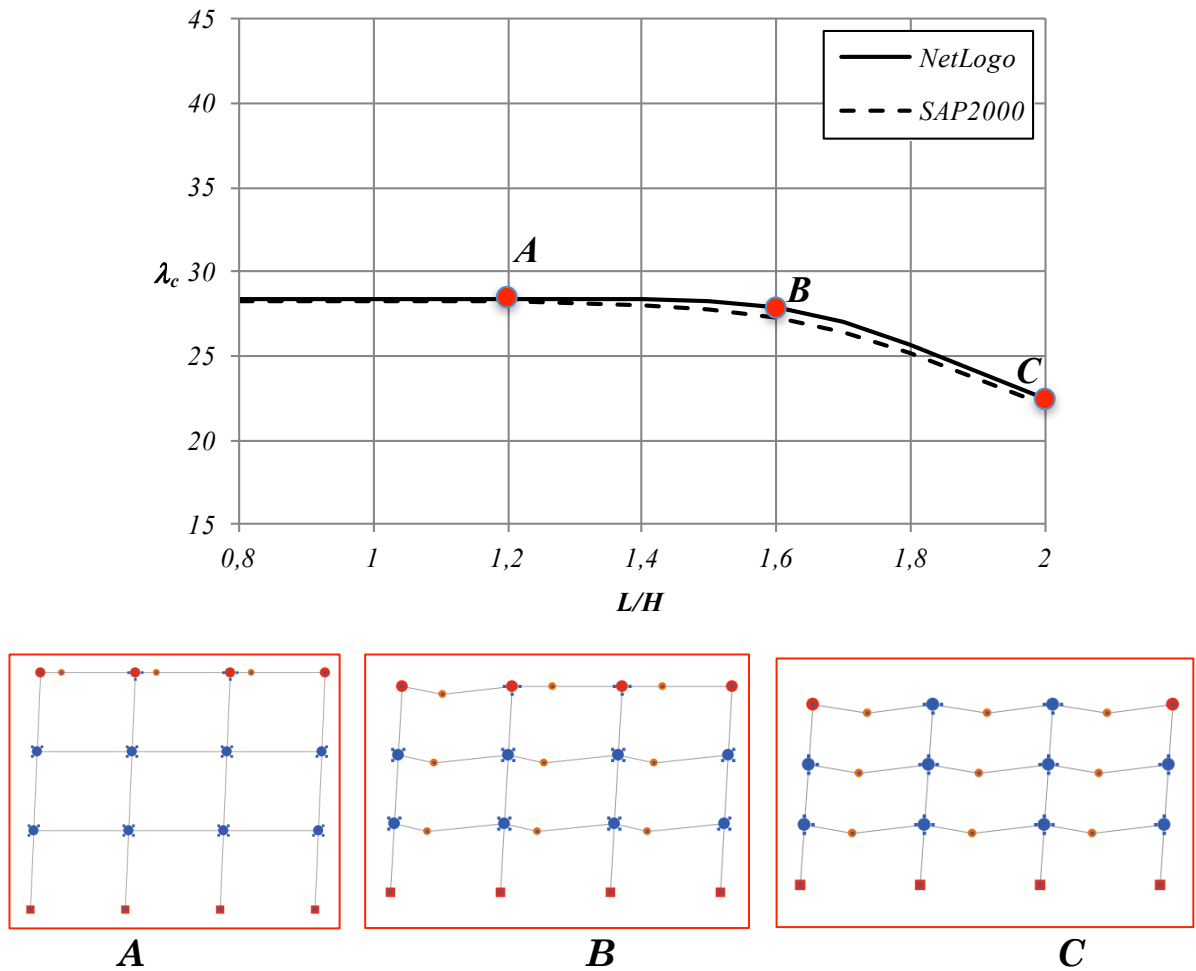

Figure 18. Collapse load multiplier vs ratio between bay length and floor height

The sensitivity of the ultimate load with respect to the ratio between the plastic moments in beams and columns has been then investigated in the range $0.3 \leq M_b/M_c \leq 1.6$ (keeping $M_c$ constant). In Figure 19 the results are reported showing again a very good agreement with the SAP2000 implementation and, as expected, that the ultimate load increases with the value of the plastic moments of the beams until the ratio $M_b/M_c = 1.4$. For higher values of the considered ratio the beams become very rigid (shear frame) and the collapse mechanism coincides with an elementary floor mechanism related to the ultimate load $\lambda_c=40$.

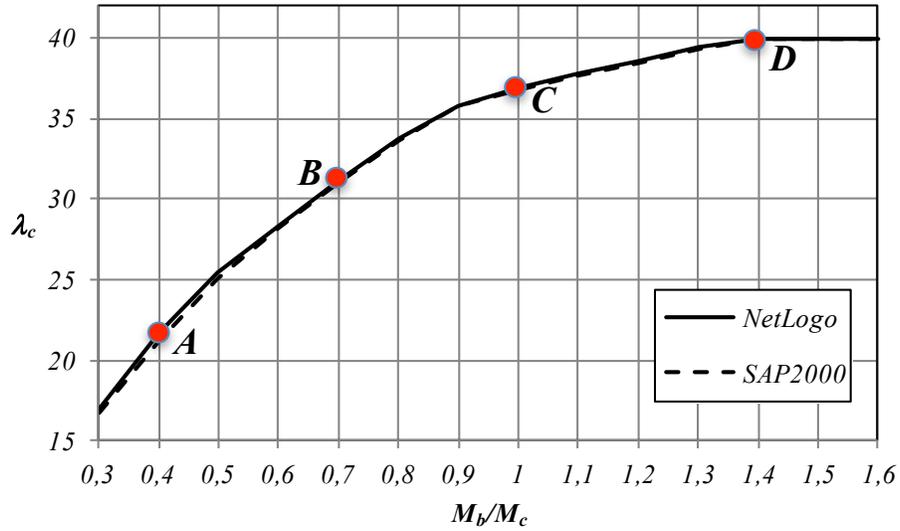

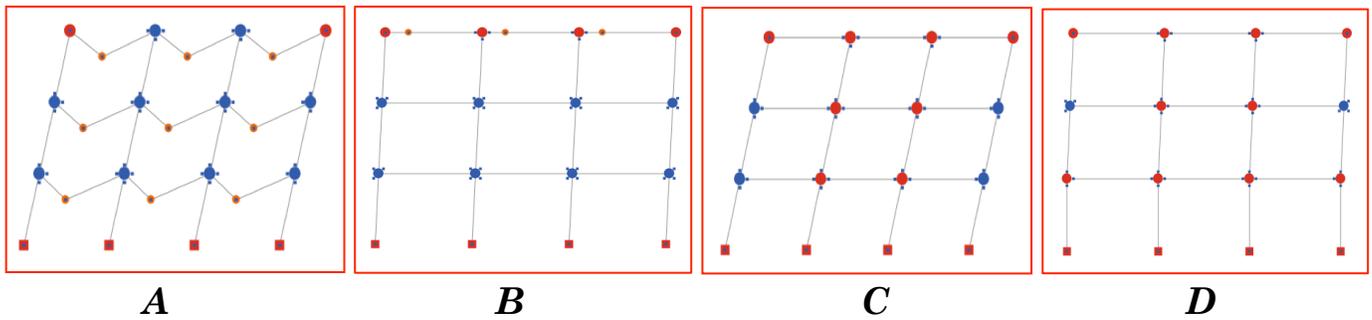

Figure 19. Collapse load multiplier vs ratio between the plastic moments of beams and columns

Finally, the last application is associated to the study of the sensitivity of the horizontal load distribution. In particular, two limit cases have been considered: one with the horizontal loads proportional to the vertical load distribution (mass proportional distribution), and one with the horizontal loads proportional to the product between the vertical load intensity and the height (inverse triangular distribution). The total vertical loads for each floor are $F_{v,1}=360$, $F_{v,2}=300$, $F_{v,3}=240$ giving an overall amount of the vertical loads equal to $\sum F_{v,i}=900$. The horizontal load distribution is progressively modulated according to a load distribution shape factor $\beta$ between the two considered limit conditions. For a given load distribution shape factor $\beta$ the horizontal force at the $i$-th storey is given by $F_{h,i}=F_{v,i}\left(1+\beta\dfrac{h_i-h_1}{h_1}\right)$ where $h_i$ represents the absolute height of the $i$-th storey and $\beta$ ranges between $0 \leq \beta \leq 1$. The horizontal loads are then normalized in such a way that their summation always corresponds to the total vertical loads (*900*), thus allowing to refer the collapse multiplier to the same value. For this reasons, the collapse multiplier corresponds in this case to the well know base shear coefficient, that is the ratio between the maximum base shear and the total weight of the structure. The results are reported in Figure 20 and show that, as the load distribution moves towards the inverse triangular distribution, as expected the ultimate carrying capacity of the frame reduces because it significantly disadvantages the higher storeys.

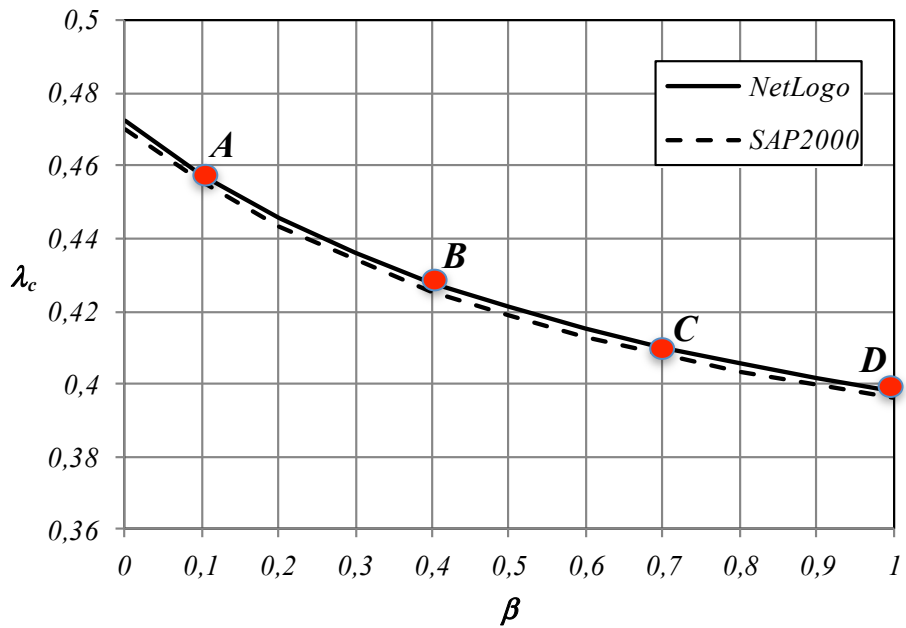

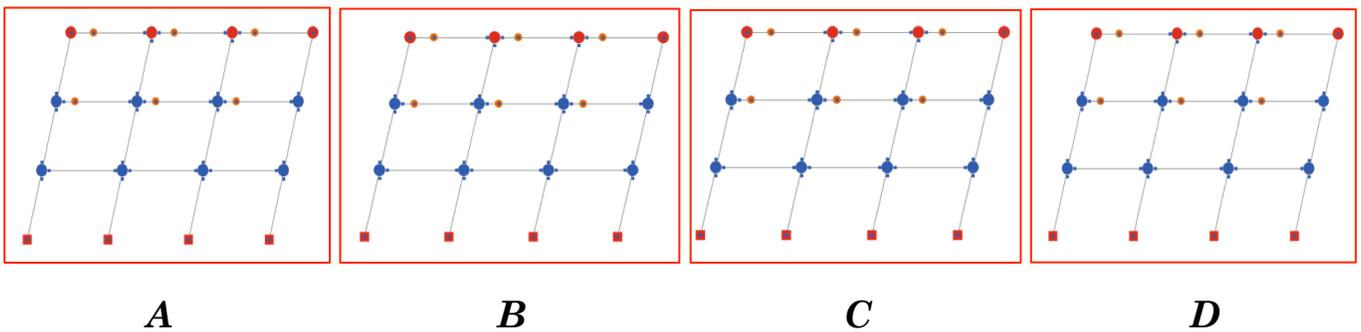

Figure 20. Collapse load multiplier vs load distribution shape factor.

# 5  Conclusions

An automatic approach for the evaluation of plastic load and failure modes of planar frames has been presented based on the generation of elementary collapse mechanisms and on their linear combination aimed at minimizing the collapse load factor.

The proposed approach makes use of an original software developed in the agent-based programming language Netlogo, which at the authors' knowledge is here employed for the first time to structural engineering. The great strength of the presented automatic approach consists in the simple visualization of each elementary and combined mechanism and on the very short computing time of their calculation.

The minimization procedure is efficiently performed by means of genetic algorithms which allow to compute an approximate collapse load factor and the related mechanism very quickly and with sufficient accuracy.

Many applications have been performed either with reference to the classical plastic analysis approach, in which all the loads increase proportionally, or with a seismic point of view considering a system of horizontal forces whose magnitude increases while the vertical loads are assumed to be constant.

The performed applications have been compared either to some of the available results provided in the literature, in the case of proportional loads, or to the results provided by non linear push over analysis in the seismic approach. A sensitivity study of the population size, of the percentage of the cross over rate and of the mutation rate in the genetic algorithm has been also performed.

The seismic applications represent an original contribution towards the limit behaviour of structures under earthquake excitations, since general trends of seismic behaviour of planar frames can be deduced from the obtained results.

An extensive parametric study has been performed aiming at evaluating the influence of some geometric and mechanical parameters, of the intensity of permanent vertical weights, and of the shape of the horizontal force distribution on the ultimate collapse load of planar frames.

The achieved results, with reference to the parametric studies, not only may provide significant information on the seismic performance of frame structures but also represent a useful tool in their optimal design.